\documentclass[aps,prl,superscriptaddress,twocolumn]{revtex4-1}

\usepackage{graphics}
\usepackage{graphicx} 
\usepackage{dcolumn}
\usepackage{bm}
\usepackage{epstopdf}
\usepackage[usenames, dvipsnames]{color}

\begin{document}


\title{Spin-gap and two-dimensional magnetic excitations in Sr$_2$IrO$_4$}



\author{S.~Calder}
\email{caldersa@ornl.gov}
\affiliation{Neutron Scattering Division, Oak Ridge National Laboratory, Oak Ridge, Tennessee 37831, USA.}

\author{D.~M.~Pajerowski}
\affiliation{Neutron Scattering Division, Oak Ridge National Laboratory, Oak Ridge, Tennessee 37831, USA.}

\author{M.~B.~Stone}
\affiliation{Neutron Scattering Division, Oak Ridge National Laboratory, Oak Ridge, Tennessee 37831, USA.}

\author{A.~F.~May}
\affiliation{Materials Science and Technology Division, Oak Ridge National Laboratory, Oak Ridge, Tennessee 37831, USA}


\begin{abstract}	
Time-of-flight inelastic neutron scattering measurements on Sr$_2$IrO$_4$ single crystals were performed to access the spin Hamiltonian in this canonical $J_{\rm eff}$=1/2 spin-orbital Mott insulator. The momentum of magnetic scattering at all inelastic energies that were measured is revealed to be $L$-independent, indicative of idealized two-dimensional in-plane correlations. By probing the in-plane energy and momentum dependence up to $\sim$80 meV we model the magnetic excitations and define a spin-gap of 0.6(1) meV. Collectively the results indicate that despite the strong spin-orbit entangled isospins an isotropic two-dimensional S=1/2 Heisenberg model Hamiltonian accurately describes the magnetic interactions, pointing to a robust analogy with unconventional superconducting cuprates. 
\end{abstract}


\maketitle

In 5$d$ oxides the paradigm of large spin-orbit coupling (SOC), appreciable Coloumb interaction ($U$) and large orbital hybridization produces strongly correlated behavior \cite{NaturePesin, doi:10.1146/annurev-conmatphys-031115-011319,doi:10.1146/annurev-conmatphys-020911-125138}. These behaviors include exotic quasiparticles such as Majorona fermions and quantum spin liquid behavior, Weyl fermions, magnetism with strong bond directionality and lattice coupling and unusual insulating states  \cite{PhysRevLett.108.127204, NatureChun, PhysRevLett.117.176603, NatureTakagi, PhysRevLett.120.237202}. The increased focus on 5$d$ materials stems from the observation that relativistic SOC drives a Mott-like insulating ground state with pseudospin $J_{\rm eff}$=1/2 magnetic moments in the iridate compound Sr$_2$IrO$_4$ \cite{PhysRevLett.101.076402, KimScience}. While the list of interesting 5$d$ compounds grows, Sr$_2$IrO$_4$ endures as a canonical material.

One surprising aspect of the physics of Sr$_2$IrO$_4$ is the similarities to the parent unconventional cuprate La$_2$CuO$_4$. The degree to which this analogy holds stands as an important outstanding question with broad implications on a wide sphere of condensed matter physics. Compelling evidence for the proximity of Sr$_2$IrO$_4$ to an unconventional superconducting state analogous to that in the cuprates was observed in the measurements of the energy and momentum dependence of magnetic excitations in Sr$_2$IrO$_4$ with resonant inelastic x-ray scattering (RIXS) \cite{PhysRevLett.108.177003}. As with the parent cuprate La$_2$CuO$_4$, an isotropic 2D Heisenberg model  describes the measured region of the excitation spectra of Sr$_2$IrO$_4$. This was in contrast to initial theoretical consideration predicting highly anisotropic (gapped) behavior \cite{PhysRevLett.102.017205}, however given the 130 meV resolution the low energy regime could not be fully accessed. Nevertheless, coupled with similarities of the ground state properties between Sr$_2$IrO$_4$ and La$_2$CuO$_4$ in terms of the layered perovskite crystal structure, antiferromagnetic ordering of pseudospin-1/2 moments, Mott insulating behavior and signatures associated with superconductivity on the surface of doped Sr$_2$IrO$_4$ provide an intriguing case  \cite{PhysRevB.49.9198,PhysRevB.87.140406,PhysRevB.87.144405,KimScience, Kim187}. This analogy has led to many open questions regarding the extent to the similarities, in particular when discussing model Hamiltonians. There is an expected strong impact of SOC on the magnetic moments of Sr$_2$IrO$_4$ ($\lambda_{\rm SO}$[Ir]$\approx$0.7 eV) while conversely being essentially negligible in La$_2$CuO$_4$ ($\lambda_{\rm SO}$[Cu]$\approx$0.01 eV). For example, this strong SOC limit in Sr$_2$IrO$_4$ is manifested in the rigid canting of the Ir moments due to the Dzyaloshinskii-Moriya (DM) interaction \cite{0953-8984-25-42-422202}. The dominant part of the magnetic Hamiltonian in the weak SOC limit is isotropic, but the introduction of SOC and Hund's coupling results in anisotropic terms \cite{PhysRevLett.102.017205}.

Experimentally, the presence or absence of an energy gap in the magnetic excitation spectra at magnetic zone center ($\pi,\pi$) delineates between isotropic or anisotropic interactions \cite{PhysRevLett.102.017205}. Resolution limitations of RIXS spectrometers have resulted in conflicting reports on the low energy excitations and whether the collective excitations in Sr$_2$IrO$_4$ differ from cuprates. While the initial measurements in Ref. \onlinecite{PhysRevLett.108.177003} did not allow access to this regime, subsequent studies with improved resolution of 30 meV have found either no indication of a spin-gap \cite{PhysRevLett.117.107001, PhysRevB.93.241102} or conversely strongly gapped excitations of the order 20-30 meV \cite{Vale_PhysRevB.92.020406, Pincini_PhysRevB.96.075162}. Separate, less direct measurements have indicated field dependent gaps of $\sim$1 meV with a cross-over from anisotropic to isotropic regimes in going from low ($H$$<$1.5 T) to high field ($H$$<$1.5 T)  \cite{PhysRevB.89.180401,PhysRevB.93.024405}. 

\begin{figure}[tb]
	\centering         
	\includegraphics[trim=0cm 9.1cm 0cm 0cm,clip=true, width=1.0\columnwidth]{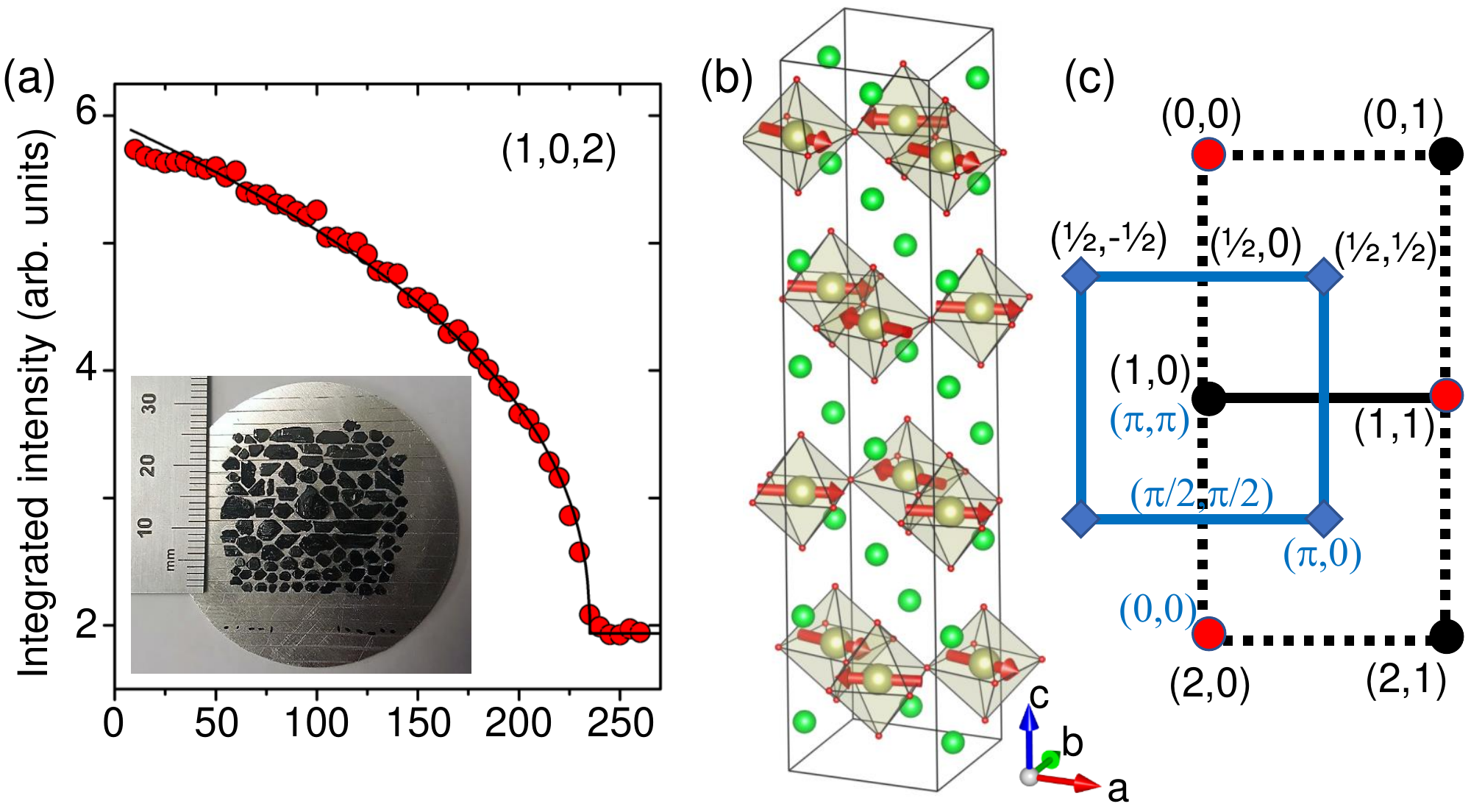}           
	\caption{\label{Figxtalarray} Sr$_2$IrO$_4$ crystal structure and magnetic ordering. (a) The single crystal array of Sr$_2$IrO$_4$ and measurement of magnetic ordering of the (1,0,2) reflection. The data (circles) are fit to a power law (line), with $\rm T_N$=240 K (b) Crystal and magnetic structure of Sr$_2$IrO$_4$. (c) Structural (red circles) and magnetic (black circles) reciprocal space. High symmetry magnetic zone boundary points are indicated by the blue diamonds. H and K points are labeled along with the labeling in the square lattice notation used to describe the dispersions. The square lattice is rotated from the conventional lattice of Sr$_2$IrO$_4$.}
\end{figure}

Here, we present time-of-flight inelastic neutron scattering (INS) measurements that directly access the low energy magnetic excitation spectra and reveal the dimensionality of the correlations. In general, INS has unique capabilities in probing magnetic excitations, with the measurements corresponding to a well understood S(Q, $\omega$) scattering cross-section. Moreover, the use of time-of-flight neutrons from spallation sources, coupled with instruments containing large detector arrays, allows ready access to four dimensional (H,K,L,E) reciprocal space. The energy resolution of neutron spectrometers additionally cover the eV down to meV energy regime, allowing the full mapping of high energy excitations as well as the inspection of low energy scattering to high precision. As such, INS has proven irreplaceable in the study of the cuprates \cite{PhysRevLett.105.247001} and measurements on Sr$_2$IrO$_4$ have been a long-standing goal. Sr$_2$IrO$_4$, however, offers technical challenges for INS measurements. Hurdles include the strong neutron absorption of iridium, small ordered moment sizes (0.2-0.3$\rm \mu_B$), rapidly falling off intensity with Q due to the Ir magnetic form factor, and small crystal sizes that together hinder the detection of magnetic signals. To overcome these issues we prepared an array of $\sim$100 single crystals of Sr$_2$IrO$_4$ with a total mass of 1.1 grams, shown in Fig.~\ref{Figxtalarray}(a). 

The largest single crystal of Sr$_2$IrO$_4$ was 300 mg, representing more than an order of magnitude increase in size compared to previous reports in the literature \cite{PhysRevB.87.140406, PhysRevB.87.144405, doi:10.1080/14786435.2015.1134835}. Crystals were grown in several batches and were found to consistently have the same ordering temperature of 240 K associated with non-deficient Sr$_2$IrO$_4$ crystals \cite{doi:10.1080/14786435.2015.1134835, PhysRevB.95.155135}. The array was aligned in the [H0L] horizontal scattering plane using a backscattering x-ray Laue and subsequent measurements with neutrons found a mosaic of ~2$^{\circ}$ FWHM. The magnetic ordering temperature of the full array was probed with the fixed elastic energy triple axis neutron spectrometer HB-1A at the High Flux Isotope Reactor (HFIR). The magnetic order of the full array was confirmed to occur at 240 K by following the magnetic (1,0,2) reflection as shown in Fig.~\ref{Figxtalarray}(a).

The INS measurements covered an energy up to $\sim$100 meV and were performed on the SEQUOIA and CNCS time-of-flight spectrometers at the Spallation Neutron Source, ORNL \cite{doi:10.1063/1.4870050}. An incident energy of 3.32 meV was utilized on CNCS to access the scattering at ($\pi$,$\pi$) and define the spin-gap. The elastic line instrumental resolution of this instrument was fit to 0.1 meV. The chosen E$\rm _i$ offers a low background since it is below the Al cut-off energy, mitigating Bragg scattering from the sample environment and Al crystal mount. On SEQUOIA, measurements were performed with incident energies of E$\rm _i=$ 20, 60 and 120 meV. On both instruments measurements were taken at fixed angles from $\psi$=$\pm30^{\circ}$, with $\psi$=$0$ corresponding to the incident neutron beam ($k_i$) being parallel to the crystallographic c-axis. This rotation range allowed coverage of a large volume of reciprocal space while minimizing neutron absorption. On SEQUOIA data were collected under the same conditions using an empty sample holder and identical Al disc with a similar mass of fomblin grease to subtract out the background scattering. All INS measurements were performed at 10 K in the ordered state.

\begin{figure}[tb]
	\centering         
	\includegraphics[trim=7.5cm 6.7cm 8.8cm 0cm,clip=true, width=1.0\columnwidth]{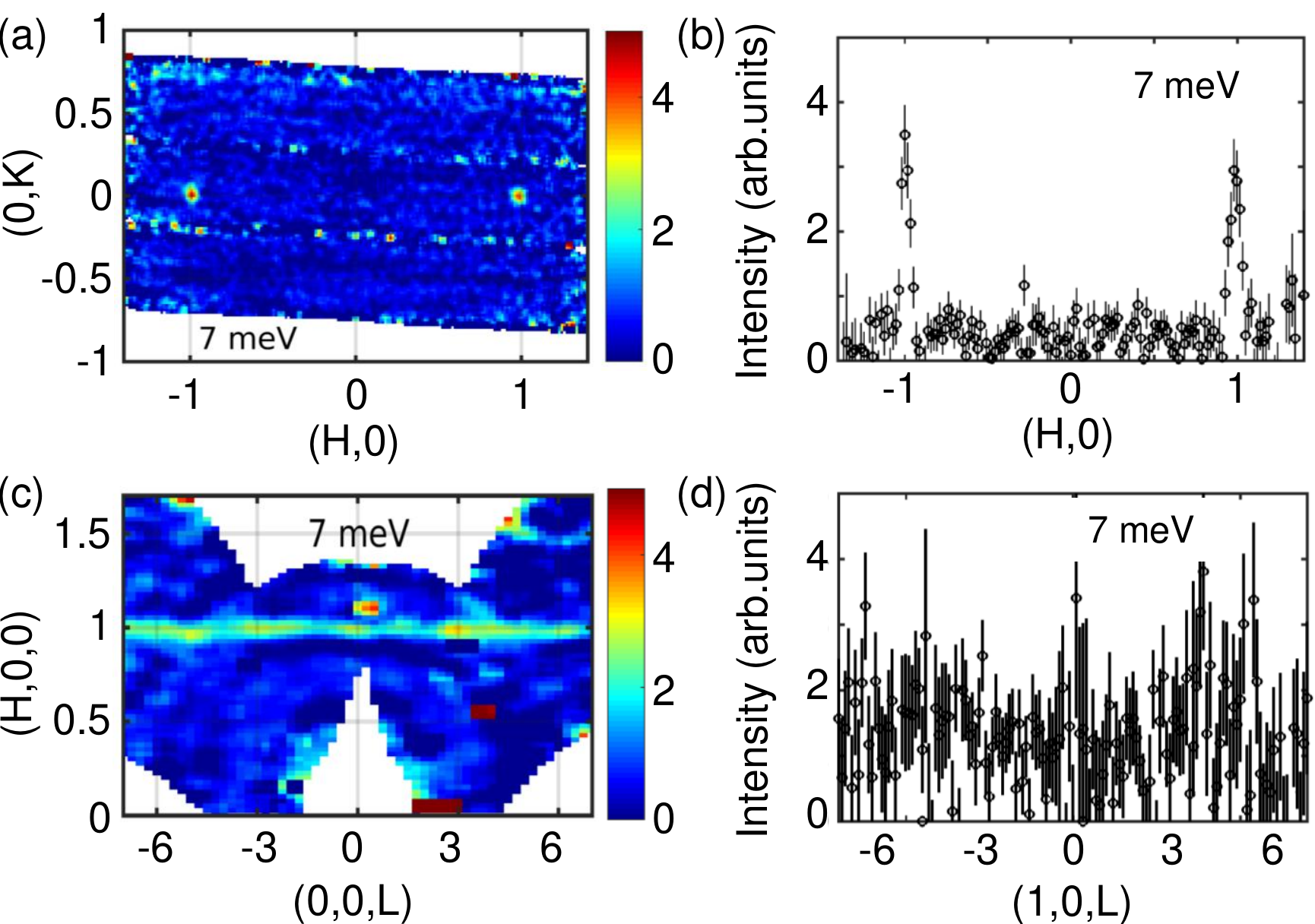} 
	\caption{\label{FigINS} Time-of-flight inelastic neutron scattering measurements of Sr$_2$IrO$_4$ on SEQUOIA. (a) The magnetic excitations around ($\pi$,$\pi$) are shown in the (H,K) plane at E=7 meV. (b) Cut along (H,0) over a K range of [-0.05,0.05] and L=[-5,5], show sharp inelastic peaks at (-1,0) and (1,0). (c) Inelastic scattering in the (H,0,L) plane at E=7 meV. (d) Cut along the rod of scattering at (1,0,L) with a K and H range of [-0.075,0.075].}
\end{figure}

The INS data show that the magnetic correlations in Sr$_2$IrO$_4$ are highly two-dimensional (2D) in nature. The antiferromagnetic order of Sr$_2$IrO$_4$ yields magnetic scattering at 10L (L=even) Bragg reflections and these elastic reflections are narrow along both H,K and L. In the INS data, sharp inelastic excitations are observed from the ($\pi,\pi$) magnetic zone center, Fig.~\ref{FigINS}(a)-(b). Conversely in the (H,0,L) plane, Fig.~\ref{FigINS}(c)-(d), the excitations are rods of scattering with no observable  momentum dependence along the $L$-direction. This $L$ behavior was observed over the fully measured inelastic energy range, providing direct experimental evidence that the magnetic correlations in Sr$_2$IrO$_4$ are 2D. This two-dimensionality has important theoretical implications in the models employed to describe the correlations. Experimental, no  $L$-dispersion offers the powerful avenue of being able to integrate the data over a large $L$-range to access wider in-plane coverage at higher statistics (signal-to-noise). This integration is particularly beneficial to time-of-flight INS that yields large volumes of (H,K,L,E) space which can then be appropriately integrated to increase access to the in-plane magnetic spectrum. It was checked that integrating over $L$ produced identical results to instead utilizing a narrow $L$ range.

\begin{figure}[tb]
	\centering         
	\includegraphics[trim=0cm 13.5cm 0cm 0cm,clip=true, width=1.0\columnwidth]{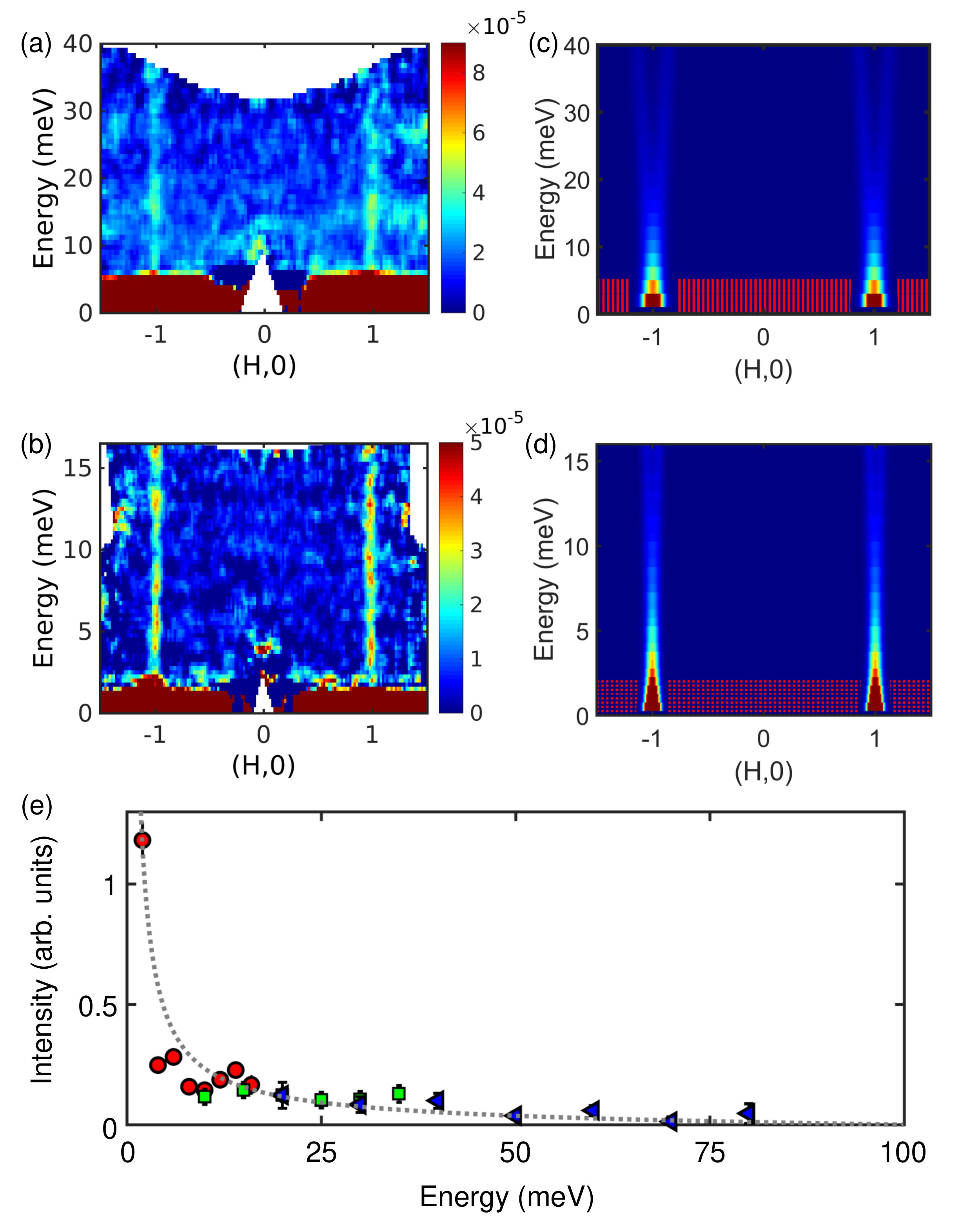} 
	\caption{\label{FigINS_model} Inelastic neutron scattering data measured on SEQUOIA for (a) E$\rm _i$=60 meV and (b) E$\rm _i$=20.5 meV. (c)-(d) These are compared with model calculations of spin wave excitations using equation 1 convoluted with the appropriate instrument energy resolution for each $\rm Ei$. (e) Extracted intensity of the inelastic scattering for E$\rm _i$=20 meV (circle), E$\rm _i$=60 meV (square), E$\rm _i$=120 meV (triangle). The intensities have been scaled to account for flux differences with the different incident energies. The dashed line is the calculated scattering intensity of the excitation as a function of energy transfer.  The largest scattering intensity occurs near the ($\pi$,$\pi$) point (0 meV) and the minimum scattering intensity is at the (0,0) wavevector and top of the excitation band (100 meV).}
\end{figure}

Intermediate incident energies, using $\rm E_i$=20.5 and 60 meV, show spin excitations consistent with RIXS, except that intensity is clearly present down to 2meV, Fig.~\ref{FigINS_model}(a)-(b). Sharp scattering is observed from ($\pi$,$\pi$) that broadens and decreases in intensity as it extends up to high energy. By inspection of Fig.~\ref{FigINS_model}(b) there is no observable spin-gap within the 2 meV resolution. Therefore, before focusing on lower energy measurements, we begin by utilizing the isotropic 2D Heisenberg model with S=1/2:

\begin{equation}
\mathcal{H}= \sum_{i,j}J_{ij}\vec{S}_i.\vec{S}_j + \Gamma S_i^zS_i^z + D(S_i^x.S_j^y - S_i^yS_j^x)
\end{equation}

\noindent with $J$ corresponding to the isotropic magnetic exchange interaction in the plane, $\Gamma$ corresponding to the symmetric exchange anisotropy and $D$ antisymmetric exchange anisotropy. The two exchange anisotropies compete, with $\Gamma$ facilitating collinear c-axis spins and $D$ promoting in-plane canting. The values for nearest neighbor Heisenberg exchange interactions have been found from previous RIXS studies of $J_1=$ 57meV,$J_2=$ -16 meV and $J_3=$ 12meV and these are utilized in the analysis of the INS data \cite{KimScience, PhysRevLett.117.107001, Pincini_PhysRevB.96.075162}. Given the lack of any observed spin-gap in the data in Fig.~\ref{FigINS_model} we use a zero-gap model.

Good agreement is seen between the measured and calculated dispersion maps in Fig~\ref{FigINS_model}(a)-(d). To further test this model, constant energy cuts of the data were taken and the scattered intensity fit to a Gaussian centered on (1,0) for E$\rm _i$=20.5 meV (energy cut range of $\pm$ 2meV), 60 meV (energy cut range of $\pm$ 5 meV) and 120 meV (energy cut range of $\pm$ 10 meV), see  Fig.~\ref{FigINS_model}(e). The different E$\rm _i$ experimental set-ups have different incident flux and so each E$\rm 
_i$ data set was normalized at overlapping inelastic energies. The INS data covered the dispersion region from ($\pi$,$\pi$) to (0,0). The calculated intensity from equation 1 closely follows the extracted intensity from the INS data, providing evidence for the applicability of an isotropic pseudo-S=1/2 2D Heisenberg model.

\begin{figure}[tb]
	\centering         
	\includegraphics[trim=0cm 0cm 0cm 0cm,clip=true, width=1.0\columnwidth]{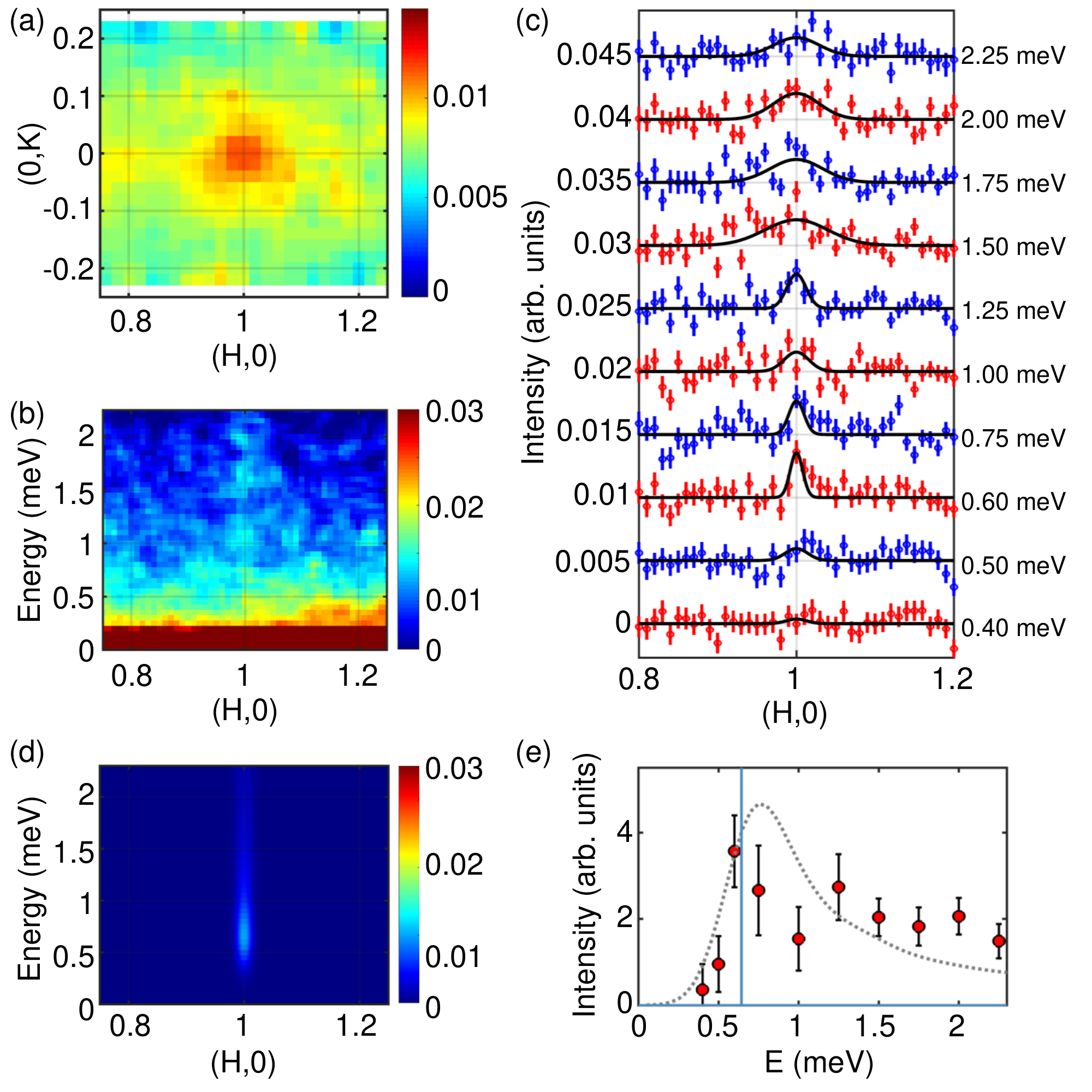} 
	\caption{\label{FigCNCS} Measurements of Sr$_2$IrO$_4$ with cold neutrons (E$\rm _i$=3.32 meV) on the CNCS. (a) Scattering centered at ($\pi$,$\pi$) is shown at 1.5 meV over an energy range of $\pm$0.25meV and L range of [-2.5,2.5]. (b) The low energy dispersion from ($\pi$,$\pi$) is shown. The K range is [-0.025,0.025] and L range is [-2.5,2.5]. (c) Cuts along (H,0) down to 0.4 meV with a K range of [-0.05,0.05], L range of [-2.5,2.5] and energy range of $\pm$0.15meV. The black lines are Gaussian fits centered on (1,0). (d) Calculated low energy scattering. (e) The extracted intensity from Gaussian fits at various energies (circles) is compared to the model intensity variation with energy, including instrument resolution, at ($\pi$,$\pi$) (dotted line). The solid line is the modeled spectra with no instrument resolution that provides a sharp peak at the spin-gap energy.}
\end{figure}

Having established the existence of magnetic scattering down to 2 meV resolution, we consider excitations emanating from the zone center ($\pi$,$\pi$), measured with a resolution of 0.1 meV, to define the spin-gap. The high resolution results are shown in Fig.~\ref{FigCNCS}. The in-plane scattering, integrated over an $L$ range of $\pm$2.5 shows a well defined scattering centered on (1,0), see Fig.~\ref{FigCNCS}(a). The measured dispersion, Fig.~\ref{FigCNCS}(b), at various energy transfers shows scattering at ($\pi$,$\pi$) well below 2 meV. To extract quantitative information, constant energy cuts, with a range of $\pm$0.15meV, along the (H,0) direction are taken and these are fit to a Gaussian peak shape centered on (1,0). The cuts and fits are shown in Fig.~\ref{FigCNCS}(c), with each energy offset by a constant factor of 5$\times10^{-3}$. The increased background in going from 2.25 meV to 0 meV, neglecting the elastic line, can be attributed to incoherent scattering from the epoxy, observable in Fig.~\ref{FigCNCS}(b). We extracted a constant fitted flat background from the data for each energy to produce the plots in Fig.~\ref{FigCNCS}(c). The corresponding intensity as a function of energy is shown in Fig.~\ref{FigCNCS}(e), with values going to zero intensity at the lowest energy indicating a finite spin-gap. To extract the spin-gap value, we then modeled the low energy scattering using equation 1 with anisotropy introduced. Using the instrument resolution the calculated intensity variation with energy was obtained. The model intensity at (1,0) was then compared to the extracted intensity from the data, with the best match shown in Fig.~\ref{FigCNCS}(e). To define the spin-gap energy the instrumental resolution was removed from the model, with the consequence of producing a sharp near delta peak at a single energy, shown by the solid line in Fig.~\ref{FigCNCS}(e). This allows a spin-gap definition of 0.6(1) meV. These results with INS therefore provide definitive evidence of the zero-field spin-gap and should serve to resolve the debate in the literate. The value of the spin-gap is well below some indications from RIXS of $\sim$20 meV \cite{Vale_PhysRevB.92.020406, Pincini_PhysRevB.96.075162}. There was an observation of an acoustic phonon mode that crossed (1,0) at 15 meV in the INS data collected here, see Fig.~\ref{FigINS_model}(a), that may explain the discrepancy from the RIXS results. Our results agree better, although still decreased from, the $\sim$1 meV from Refs.~\onlinecite{PhysRevB.89.180401,PhysRevB.93.024405}. 

From the INS results, Sr$_2$IrO$_4$ can be placed in the weak to intermediate SOC limit. Given the large magnetic excitation bandwidth of 200 meV the spin-gap is finite but negligible. For comparison the measured spin-gap in La$_2$CuO$_4$ from INS is 5 meV, with an excitation bandwidth of 300 meV, despite the order of magnitude reduced SOC in Cu compared to Ir. This is surprising considering the importance of SOC in generating the electronic ground state, however the results indicate a dramatically reduced impact of SOC on the manifested magnetic correlations. The Ir$^{4+}$ ion has $\lambda_{\rm SO}$=0.7 eV in Sr$_2$IrO$_4$ and the crystal field splitting is $\Delta_{\rm oct}$=3.5 eV, placing Sr$_2$IrO$_4$ in the intermediate coupling limit. With the tetragonal distortion of the oxygen coordination sphere, the first excited state in the strong crystal field picture, as found in cuprates, then is strongly mixed with the ground state and gives an easy plane anisotropy that makes the system quasi-2D and is the manifestation of the large SOC. Then there exists a slight rhombic term in the plane, along the $b$-axis \cite{PhysRevB.96.155102}, that produces the small but finite measured gap. The small gap, and therefore small anisotropy, coupled with the measurement of 2D correlations in Fig.~\ref{FigINS} indicates an isotropic 2D model will robustly describe magnetic correlations in Sr$_2$IrO$_4$. 

Collectively the results presented provide compelling evidence for the mapping of the physics of Sr$_2$IrO$_4$ onto the parent cuprate La$_2$CuO$_4$. While the high energy spin excitations have been followed with RIXS, the scattering cross-section measured is not fully described. Therefore the similarities of the RIXS data and the INS measurements are in some respects remarkable. These results in themselves provide a system independent verification of the quantitative data available from RIXS at high energies and show the power of combining results from RIXS and INS. In this case the limitation of RIXS is the energy resolution, although strong advances have been made \cite{JunghoUHResRIXS}. INS, therefore, offers the unparalleled ability to probe low energy sub-meV signals over the full reciprocal space. This has allowed the definition of the small spin-gap energy in Sr$_2$IrO$_4$, resolving contradictory reports from various techniques.

In conclusion, the spin-gap of Sr$_2$IrO$_4$ is measured to be 0.6(1) meV and the magnetic correlations shown to be highly two-dimensional. These results were obtained through inelastic neutron scattering measurements performed on single crystals of Sr$_2$IrO$_4$ within the magnetically order phase. Well-defined excitations were revealed using high resolution measurements to give a spin-gap value. The excitation spectrum is found to be strongly two-dimensional with no measurable out-of-plane $L$-dispersion. Collectively the INS measurements show that an isotropic S=1/2 2D Heisenberg model is applicable for the entire magnetic excitation regime of Sr$_2$IrO$_4$. The implications reinforce the analogy with cuprates despite the presence of strong SOC on the Ir ion that would be expected to result in anisotropic behavior.

\begin{acknowledgments}
This research used resources at the High Flux Isotope Reactor and Spallation Neutron Source, a DOE Office of Science User Facility operated by the Oak Ridge National Laboratory. This work was supported by the U. S. Department of
Energy, Office of Science, Basic Energy Sciences, Materials Sciences and Engineering Division.
\end{acknowledgments}


%

\end{document}